\input harvmac
%\draftmode

%%%%%%%%%%%%%%%%%%%
%%%%%% Fonts %%%%%%
%%%%%%%%%%%%%%%%%%%

\newfam\black % for fields like Z,R,C
\font\blackboard=msbm10
\font\blackboards=msbm7
\font\blackboardss=msbm5
\textfont\black=\blackboard
\scriptfont\black=\blackboards
\scriptscriptfont\black=\blackboardss
\def\Bbb#1{{\fam\black\relax#1}}

%%%%%%%%%%%%%%%%%%%%%%%%%%
%%%       Macros       %%%
%%%%%%%%%%%%%%%%%%%%%%%%%%

\def\subsubsec#1{\ifnum\lastpenalty>9000\else\bigbreak\fi
\noindent{\it{#1}}\par\nobreak\medskip\nobreak}

\def\QR{\Bbb{R}}
\def\QX{\Bbb{X}}
\def\QZ{\Bbb{Z}}
\def\CN{{\cal N}}
\def\CJ{{\cal J}}
\def\CL{{\cal L}}
\def\CH{{\cal H}}

\def\O{{\cal O}}

\

\def\p{\partial}
\def\id{{\bf 1}}
\def\tPi{{\tilde \Pi}}

\def\tX{{\tilde X}}
\def\ty{{\tilde y}}
\def\tz{{\tilde z}}
\def\frac#1#2{{#1 \over #2}}

%%%%%%%%%%%%%%%%%%%%%%%%%%%%%%
%%%       References       %%%
%%%%%%%%%%%%%%%%%%%%%%%%%%%%%%

% Nongeometry for dummies

\lref\HullIN{
  C.~M.~Hull, ``A geometry for non-geometric string backgrounds,''
  JHEP {\bf 0510}, 065 (2005) [arXiv:hep-th/0406102].
%%CITATION = HEP-TH 0406102;%%
}
\lref\FlournoyVN{
  A.~Flournoy, B.~Wecht and B.~Williams, ``Constructing nongeometric
  vacua in string theory,'' Nucl.\ Phys.\ B {\bf 706}, 127 (2005)
  [arXiv:hep-th/0404217].
%%CITATION = HEP-TH 0404217;%%
}
\lref\FlournoyXE{
  A.~Flournoy and B.~Williams, ``Nongeometry, duality twists, and the
  worldsheet,'' arXiv:hep-th/0511126.
%%CITATION = HEP-TH 0511126;%%
}
\lref\HellermanAX{
  S.~Hellerman, J.~McGreevy and B.~Williams, ``Geometric constructions
  of nongeometric string theories,'' JHEP {\bf 0401}, 024 (2004)
  [arXiv:hep-th/0208174].
%%CITATION = HEP-TH 0208174;%%
}
\lref\SheltonCF{
  J.~Shelton, W.~Taylor and B.~Wecht, ``Nongeometric flux
  compactifications,'' JHEP {\bf 0510}, 085 (2005)
  [arXiv:hep-th/0508133].
%%CITATION = HEP-TH 0508133;%%
}

% Mathy nongeometry

\lref\BouwknegtVB{
  P.~Bouwknegt, J.~Evslin and V.~Mathai, ``T-duality: Topology change
  from $H$-flux,'' Commun.\ Math.\ Phys.\ {\bf 249}, 383 (2004)
  [arXiv:hep-th/0306062].
%%CITATION = HEP-TH 0306062;%%
}
\lref\BouwknegtWP{
  P.~Bouwknegt, J.~Evslin and V.~Mathai, ``On the topology and $H$-flux
  of T-dual manifolds,'' Phys.\ Rev.\ Lett.\ {\bf 92}, 181601 (2004)
  [arXiv:hep-th/0312052].
%%CITATION = HEP-TH 0312052;%%
}
\lref\MathaiQQ{
  V.~Mathai and J.~M.~Rosenberg, ``T-duality for torus bundles via
  noncommutative topology,'' Commun.\ Math.\ Phys.\ {\bf 253}, 705
  (2004) [arXiv:hep-th/0401168].
%%CITATION = HEP-TH 0401168;%%
}
\lref\MathaiQC{
  V.~Mathai and J.~M.~Rosenberg, ``On mysteriously missing T-duals,
  $H$-flux and the T-duality group,'' arXiv:hep-th/0409073.
%%CITATION = HEP-TH 0409073;%%
}
\lref\BouwknegtAP{
  P.~Bouwknegt, K.~Hannabuss and V.~Mathai, ``Nonassociative tori and
  applications to T-duality,'' arXiv:hep-th/0412092.
%%CITATION = HEP-TH 0412092;%%
}
\lref\MathaiFD{
  V.~Mathai and J.~Rosenberg, ``T-duality for torus bundles with
  $H$-fluxes via noncommutative topology. II: The high-dimensional case
  and the T-duality group,'' arXiv:hep-th/0508084.
%%CITATION = HEP-TH 0508084;%%
}

% Asymmetric orbifolds

\lref\AokiSM{
  K.~Aoki, E.~D'Hoker and D.~H.~Phong, ``On the construction of
  asymmetric orbifold models,'' Nucl.\ Phys.\ B {\bf 695}, 132 (2004)
  [arXiv:hep-th/0402134].
%%CITATION = HEP-TH 0402134;%%
}

% Double your pleasure, double your fun

\lref\CremmerCT{
  E.~Cremmer, B.~Julia, H.~Lu and C.~N.~Pope, ``Dualisation of
  dualities. I,'' Nucl.\ Phys.\ B {\bf 523}, 73 (1998)
  [arXiv:hep-th/9710119].
%%CITATION = HEP-TH 9710119;%%
}
\lref\MaharanaMY{
  J.~Maharana and J.~H.~Schwarz, ``Noncompact symmetries in string
  theory,'' Nucl.\ Phys.\ B {\bf 390}, 3 (1993)
  [arXiv:hep-th/9207016].
%%CITATION = HEP-TH 9207016;%%
}
\lref\DuffTF{
  M.~J.~Duff, ``Duality Rotations In String Theory,'' Nucl.\ Phys.\ B
  {\bf 335}, 610 (1990).
%%CITATION = NUPHA,B335,610;%%
}
\lref\TseytlinNB{
  A.~A.~Tseytlin, ``Duality Symmetric Formulation Of String World
  Sheet Dynamics,'' Phys.\ Lett.\ B {\bf 242}, 163 (1990).
%%CITATION = PHLTA,B242,163;%%
}
\lref\TseytlinVA{
  A.~A.~Tseytlin, ``Duality Symmetric Closed String Theory And
  Interacting Chiral Scalars,'' Nucl.\ Phys.\ B {\bf 350}, 395 (1991).
%%CITATION = NUPHA,B350,395;%%
}

% Generalized Calabi-Yauism

\lref\HitchinUT{
  N.~Hitchin, ``Generalized Calabi-Yau manifolds,'' Quart.\ J.\ Math.\
  Oxford Ser.\ {\bf 54}, 281 (2003) [arXiv:math.dg/0209099].
%%CITATION = MATH-DG 0209099;%%
}
\lref\GualtieriThesis{
 M.~Gualtieri, ``Generalized complex geometry", arXiv:math.DG/0401221.
 }
\lref\GrangeNM{
  P.~Grange and R.~Minasian, ``Tachyon condensation and D-branes in
  generalized geometries,'' arXiv:hep-th/0512185.
%%CITATION = HEP-TH 0512185;%%
}

% Connected vacua

\lref\StromingerCZ{
  A.~Strominger, ``Massless black holes and conifolds in string
  theory,'' Nucl.\ Phys.\ B {\bf 451}, 96 (1995)
  [arXiv:hep-th/9504090].
%%CITATION = HEP-TH 9504090;%%
}
\lref\GreeneHU{
  B.~R.~Greene, D.~R.~Morrison and A.~Strominger, ``Black hole
  condensation and the unification of string vacua,'' Nucl.\ Phys.\ B
  {\bf 451}, 109 (1995) [arXiv:hep-th/9504145].
%%CITATION = HEP-TH 9504145;%%
}

% all is in flux
\lref\GukovYA{
  S.~Gukov, C.~Vafa and E.~Witten, ``CFT's from Calabi-Yau
  four-folds,'' Nucl.\ Phys.\ B {\bf 584}, 69 (2000) [Erratum-ibid.\ B
  {\bf 608}, 477 (2001)] [arXiv:hep-th/9906070].
%%CITATION = HEP-TH 9906070;%%
}
\lref\TaylorII{
  T.~R.~Taylor and C.~Vafa, ``RR flux on Calabi-Yau and partial
  supersymmetry breaking,'' Phys.\ Lett.\ B {\bf 474}, 130 (2000)
  [arXiv:hep-th/9912152].
%%CITATION = HEP-TH 9912152;%%
}
\lref\VafaWI{
  C.~Vafa,
  ``Superstrings and topological strings at large $N$,''
  J.\ Math.\ Phys.\  {\bf 42}, 2798 (2001)
  [arXiv:hep-th/0008142].
%%CITATION = HEP-TH 0008142;%%
}
\lref\LawrenceZK{
  A.~Lawrence and J.~McGreevy, ``Local string models of soft
  supersymmetry breaking,'' JHEP {\bf 0406}, 007 (2004)
  [arXiv:hep-th/0401034].
%%CITATION = HEP-TH 0401034;%%
}
\lref\LawrenceKJ{
  A.~Lawrence and J.~McGreevy, ``Remarks on branes, fluxes, and soft
  SUSY breaking,'' arXiv:hep-th/0401233.
%%CITATION = HEP-TH 0401233;%%
}
\lref\KachruJR{
  S.~Kachru and A.~K.~Kashani-Poor, ``Moduli potentials in type IIA
  compactifications with RR and NS flux,'' JHEP {\bf 0503}, 066 (2005)
  [arXiv:hep-th/0411279].
%%CITATION = HEP-TH 0411279;%%
}

% Heterotic Beckerism

\lref\BeckerYV{
  K.~Becker, M.~Becker, K.~Dasgupta and P.~S.~Green,
  ``Compactifications of heterotic theory on non-Kaehler complex
  manifolds.  I,'' JHEP {\bf 0304}, 007 (2003) [arXiv:hep-th/0301161].
%%CITATION = HEP-TH 0301161;%%
}
\lref\BeckerGQ{
  K.~Becker, M.~Becker, K.~Dasgupta and S.~Prokushkin, ``Properties of
  heterotic vacua from superpotentials,'' Nucl.\ Phys.\ B {\bf 666},
  144 (2003) [arXiv:hep-th/0304001].
%%CITATION = HEP-TH 0304001;%%
}
\lref\BeckerSH{
  K.~Becker, M.~Becker, P.~S.~Green, K.~Dasgupta and E.~Sharpe,
  ``Compactifications of heterotic strings on non-Kaehler complex
  manifolds.  II,'' Nucl.\ Phys.\ B {\bf 678}, 19 (2004)
  [arXiv:hep-th/0310058].
%%CITATION = HEP-TH 0310058;%%
}

% Mirror symmetry

\lref\homologicalmirror{
 M.~Kontsevich, ``Homological Algebra of Mirror Symmetry", in
 Proceedings of the International Congress of Mathematicians, pages
 120Ð139, Birkh\"auser, 1995, arXiv:alg-geom/9411018.
}
\lref\StromingerIT{
  A.~Strominger, S.~T.~Yau and E.~Zaslow, ``Mirror symmetry is
  T-duality,'' Nucl.\ Phys.\ B {\bf 479}, 243 (1996)
  [arXiv:hep-th/9606040].
%%CITATION = HEP-TH 9606040;%%
}

%%Fun with fluxes
\lref\KachruSK{
  S.~Kachru, M.~B.~Schulz, P.~K.~Tripathy and S.~P.~Trivedi, ``New
  supersymmetric string compactifications,'' JHEP {\bf 0303}, 061
  (2003) [arXiv:hep-th/0211182].
%%CITATION = HEP-TH 0211182;%%
}
\lref\KachruHE{
  S.~Kachru, M.~B.~Schulz and S.~Trivedi, ``Moduli stabilization from
  fluxes in a simple IIB orientifold,'' JHEP {\bf 0310}, 007 (2003)
  [arXiv:hep-th/0201028].
%%CITATION = HEP-TH 0201028;%%
}
\lref\FreyHF{
  A.~R.~Frey and J.~Polchinski, ``$N = 3$ warped compactifications,''
  Phys.\ Rev.\ D {\bf 65}, 126009 (2002) [arXiv:hep-th/0201029].
%%CITATION = HEP-TH 0201029;%%
}
\lref\TripathyQW{
  P.~K.~Tripathy and S.~P.~Trivedi, ``Compactification with flux on K3
  and tori,'' JHEP {\bf 0303}, 028 (2003) [arXiv:hep-th/0301139].
%%CITATION = HEP-TH 0301139;%%
}
\lref\DeWolfeNS{
  O.~DeWolfe, A.~Giryavets, S.~Kachru and W.~Taylor, ``Enumerating
  flux vacua with enhanced symmetries,'' JHEP {\bf 0502}, 037 (2005)
  [arXiv:hep-th/0411061].
%%CITATION = HEP-TH 0411061;%%
}
\lref\DeWolfeUU{
  O.~DeWolfe, A.~Giryavets, S.~Kachru and W.~Taylor, ``Type IIA moduli
  stabilization,'' JHEP {\bf 0507}, 066 (2005) [arXiv:hep-th/0505160].
%%CITATION = HEP-TH 0505160;%%
}
\lref\GrimmUA{
  T.~W.~Grimm and J.~Louis, ``The effective action of type IIA
  Calabi-Yau orientifolds,'' Nucl.\ Phys.\ B {\bf 718}, 153 (2005)
  [arXiv:hep-th/0412277].
%%CITATION = HEP-TH 0412277;%%
}
\lref\CamaraDC{
  P.~G.~C\'amara, A.~Font and L.~E.~Ib\'a\~nez,
  ``Fluxes, moduli fixing and MSSM-like vacua in a simple IIA orientifold,''
  JHEP {\bf 0509}, 013 (2005)
  [arXiv:hep-th/0506066].
%%CITATION = HEP-TH 0506066;%%
}
\lref\LoweQY{
  D.~A.~Lowe, H.~Nastase and S.~Ramgoolam, ``Massive IIA string theory
  and matrix theory compactification,'' Nucl.\ Phys.\ B {\bf 667}, 55
  (2003) [arXiv:hep-th/0303173].
%%CITATION = HEP-TH 0303173;%%
}
\lref\KaloperYR{
  N.~Kaloper and R.~C.~Myers, ``The $O(d,d)$ story of massive
  supergravity,'' JHEP {\bf 9905}, 010 (1999) [arXiv:hep-th/9901045].
%%CITATION = HEP-TH 9901045;%%
}

% c=1 boundary states

\lref\Friedannotes{
  D.~Friedan, ``The Space of Conformal Boundary Conditions for the
  $c=1$ Gaussian Model", 1999 (unpublished note); and ``The Space of
  Conformal Boundary Conditions for the $c=1$ Gaussian Model (more)",
  2003 (unpublished note); avaliable at {\tt
  http://www.physics.rutgers.edu/$\sim$friedan/ }
}
\lref\TsengAX{
  L.~S.~Tseng, ``A note on $c = 1$ Virasoro boundary states and
 asymmetric shift orbifolds,'' JHEP {\bf 0204}, 051 (2002)
 [arXiv:hep-th/0201254].
%%CITATION = HEP-TH 0201254;%%
}
\lref\GaberdielZQ{
  M.~R.~Gaberdiel and A.~Recknagel, ``Conformal boundary states for
  free bosons and fermions,'' JHEP {\bf 0111}, 016 (2001)
  [arXiv:hep-th/0108238].
%%CITATION = HEP-TH 0108238;%%
}
\lref\JanikHB{
  R.~A.~Janik, ``Exceptional boundary states at $c = 1$,'' Nucl.\ Phys.\
  B {\bf 618}, 675 (2001) [arXiv:hep-th/0109021].
%%CITATION = HEP-TH 0109021;%%
}
\lref\CappelliWQ{
  A.~Cappelli and G.~D'Appollonio, ``Boundary states of $c = 1$ and $3/2$
  rational conformal field theories,'' JHEP {\bf 0202}, 039 (2002)
  [arXiv:hep-th/0201173].
%%CITATION = HEP-TH 0201173;%%
}
\lref\BuscherSK{
  T.~H.~Buscher, ``A Symmetry Of The String Background Field
  Equations,'' Phys.\ Lett.\ B {\bf 194}, 59 (1987).
%%CITATION = PHLTA,B194,59;%%
}
\lref\GiveonRev{
  A.~Giveon, M.~Porrati and E.~Rabinovici, ``Target space duality in
  string theory,'' Phys. Rept. 244, 77 (1994) [arXiv:hep-th/9401139].
%%CITATION = HEP-TH 9401139;%%
  }
\lref\DabholkarSY{
  A.~Dabholkar and C.~Hull, ``Duality twists, orbifolds, and fluxes,''
  JHEP {\bf 0309}, 054 (2003) [arXiv:hep-th/0210209].
%%CITATION = HEP-TH 0210209;%%
}
\lref\DabholkarVE{
  A.~Dabholkar and C.~Hull, ``Generalised T-duality and non-geometric
  backgrounds,'' arXiv:hep-th/0512005.
%%CITATION = HEP-TH 0512005;%%
}
\lref\HullHK{
  C.~M.~Hull and R.~A.~Reid-Edwards, ``Flux compactifications of
  string theory on twisted tori,'' arXiv:hep-th/0503114.
%%CITATION = HEP-TH 0503114;%%
}
\lref\VilladoroCU{
  G.~Villadoro and F.~Zwirner, ``$N = 1$ effective potential from dual
  type-IIA D6/O6 orientifolds with general fluxes,'' JHEP {\bf 0506},
  047 (2005) [arXiv:hep-th/0503169].
%%CITATION = HEP-TH 0503169;%%
}
\lref\DerendingerJN{
  J.~P.~Derendinger, C.~Kounnas, P.~M.~Petropoulos and F.~Zwirner,
  ``Superpotentials in IIA compactifications with general fluxes,''
  Nucl.\ Phys.\ B {\bf 715}, 211 (2005) [arXiv:hep-th/0411276].
%%CITATION = HEP-TH 0411276;%%
}
\lref\ScherkZR{
  J.~Scherk and J.~H.~Schwarz, ``How To Get Masses From Extra
  Dimensions,'' Nucl.\ Phys.\ B {\bf 153}, 61 (1979).
%%CITATION = NUPHA,B153,61;%%
}
\lref\GrayEA{
  J.~Gray and E.~Hackett-Jones, ``On T-folds, G-structures and
  supersymmetry,'' arXiv:hep-th/0506092.
%%CITATION = HEP-TH 0506092;%%
}
\lref\DallAgataFF{
  G.~Dall'Agata and S.~Ferrara, ``Gauged supergravity algebras from
  twisted tori compactifications with fluxes,'' Nucl.\ Phys.\ B {\bf
  717}, 223 (2005) [arXiv:hep-th/0502066].
%%CITATION = HEP-TH 0502066;%%
}
\lref\GranaSN{
  M.~Gra\~na, R.~Minasian, M.~Petrini and A.~Tomasiello, ``Generalized
  structures of N = 1 vacua,'' JHEP {\bf 0511}, 020 (2005)
  [arXiv:hep-th/0505212].
%%CITATION = HEP-TH 0505212;%%
}
\lref\GranaSV{
  M.~Gra\~na, R.~Minasian, M.~Petrini and A.~Tomasiello, ``Type II
  strings and generalized Calabi-Yau manifolds,'' Comptes Rendus
  Physique {\bf 5}, 979 (2004) [arXiv:hep-th/0409176].
%%CITATION = HEP-TH 0409176;%%
}
%\GranaBG
\lref\GranaBG{
  M.~Gra\~na, R.~Minasian, M.~Petrini and A.~Tomasiello,
  ``Supersymmetric backgrounds from generalized Calabi-Yau
  manifolds,'' JHEP {\bf 0408}, 046 (2004) [arXiv:hep-th/0406137].
%%CITATION = HEP-TH 0406137;%%
}
\lref\GurrieriIW{
  S.~Gurrieri and A.~Micu, ``Type IIB theory on half-flat manifolds,''
  Class.\ Quant.\ Grav.\ {\bf 20}, 2181 (2003) [arXiv:hep-th/0212278].
%%CITATION = HEP-TH 0212278;%%
}
\lref\GurrieriDT{
  S.~Gurrieri, A.~Lukas and A.~Micu, ``Heterotic on half-flat,''
  Phys.\ Rev.\ D {\bf 70}, 126009 (2004) [arXiv:hep-th/0408121].
%%CITATION = HEP-TH 0408121;%%
}
\lref\FidanzaZI{
  S.~Fidanza, R.~Minasian and A.~Tomasiello, ``Mirror symmetric
  $SU(3)$-structure manifolds with NS fluxes,'' Commun.\ Math.\ Phys.\
  {\bf 254}, 401 (2005) [arXiv:hep-th/0311122].
%%CITATION = HEP-TH 0311122;%%
}
\lref\CardosoHD{
  G.~L.~Cardoso, G.~Curio, G.~Dall'Agata, D.~L\"ust, P.~Manousselis and
  G.~Zoupanos, ``Non-K\"ahler string backgrounds and their five torsion
  classes,'' Nucl.\ Phys.\ B {\bf 652}, 5 (2003)
  [arXiv:hep-th/0211118].
%%CITATION = HEP-TH 0211118;%%
}
\lref\CardosoAF{
  G.~L.~Cardoso, G.~Curio, G.~Dall'Agata and D.~L\"ust, ``BPS action and
  superpotential for heterotic string compactifications with fluxes,''
  JHEP {\bf 0310}, 004 (2003) [arXiv:hep-th/0306088].
%%CITATION = HEP-TH 0306088;%%
}
\lref\SeroneSV{
  M.~Serone and M.~Trapletti, ``String vacua with flux from
  freely-acting orbifolds,'' JHEP {\bf 0401}, 012 (2004)
  [arXiv:hep-th/0310245].
%%CITATION = HEP-TH 0310245;%%
}
\lref\BeckerII{
  K.~Becker, M.~Becker, K.~Dasgupta and R.~Tatar, ``Geometric
  transitions, non-K\"ahler geometries and string vacua,'' Int.\ J.\
  Mod.\ Phys.\ A {\bf 20}, 3442 (2005) [arXiv:hep-th/0411039].
%%CITATION = HEP-TH 0411039;%%
}
\lref\FreyRN{
  A.~R.~Frey, ``Notes on $SU(3)$ structures in type IIB supergravity,''
  JHEP {\bf 0406}, 027 (2004) [arXiv:hep-th/0404107].
%%CITATION = HEP-TH 0404107;%%
}
\lref\deCarlosKH{
  B.~de Carlos, S.~Gurrieri, A.~Lukas and A.~Micu, ``Moduli
  stabilisation in heterotic string compactifications,''
  arXiv:hep-th/0507173.
%%CITATION = HEP-TH 0507173;%%
}
\lref\AlexanderEQ{
  S.~Alexander, K.~Becker, M.~Becker, K.~Dasgupta, A.~Knauf and
  R.~Tatar, ``In the realm of the geometric transitions,'' Nucl.\
  Phys.\ B {\bf 704}, 231 (2005) [arXiv:hep-th/0408192].
%%CITATION = HEP-TH 0408192;%%
}
\lref\MicuTZ{
  A.~Micu, ``Heterotic compactifications and nearly-Kaehler
  manifolds,'' Phys.\ Rev.\ D {\bf 70}, 126002 (2004)
  [arXiv:hep-th/0409008].
%%CITATION = HEP-TH 0409008;%%
}
\lref\ChuangQD{
  W.~y.~Chuang, S.~Kachru and A.~Tomasiello, ``Complex / symplectic
  mirrors,'' arXiv:hep-th/0510042.
%%CITATION = HEP-TH 0510042;%%
}
\lref\TomasielloBP{
  A.~Tomasiello, ``Topological mirror symmetry with fluxes,'' JHEP
  {\bf 0506}, 067 (2005) [arXiv:hep-th/0502148].
%%CITATION = HEP-TH 0502148;%%
}

\lref\GurrieriWZ{
  S.~Gurrieri, J.~Louis, A.~Micu and D.~Waldram, ``Mirror symmetry in
  generalized Calabi-Yau compactifications,'' Nucl.\ Phys.\ B {\bf
  654}, 61 (2003) [arXiv:hep-th/0211102].
%%CITATION = HEP-TH 0211102;%%
}
\lref\GranaNY{
  M.~Gra\~na, J.~Louis and D.~Waldram, ``Hitchin functionals in $N = 2$
  supergravity,'' arXiv:hep-th/0505264.
%%CITATION = HEP-TH 0505264;%%
}
\lref\HouseYC{
  T.~House and E.~Palti,
  ``Effective action of (massive) IIA on manifolds with $SU(3)$ structure,''
  Phys.\ Rev.\ D {\bf 72}, 026004 (2005)
  [arXiv:hep-th/0505177].
%%CITATION = HEP-TH 0505177;%%
}
\lref\FidanzaZI{
  S.~Fidanza, R.~Minasian and A.~Tomasiello, ``Mirror symmetric
  $SU(3)$-structure manifolds with NS fluxes,'' Commun.\ Math.\ Phys.\
  {\bf 254}, 401 (2005) [arXiv:hep-th/0311122].
%%CITATION = HEP-TH 0311122;%%
}
%\LindstromIW
\lref\LindstromIW{
  U.~Lindstrom, R.~Minasian, A.~Tomasiello and M.~Zabzine,
  ``Generalized complex manifolds and supersymmetry,'' Commun.\ Math.\
  Phys.\ {\bf 257}, 235 (2005) [arXiv:hep-th/0405085].
%%CITATION = HEP-TH 0405085;%%
}
%\LindstromEH
\lref\LindstromEH{
  U.~Lindstrom, ``Generalized N = (2,2) supersymmetric non-linear
  sigma models,'' Phys.\ Lett.\ B {\bf 587}, 216 (2004)
  [arXiv:hep-th/0401100].
%%CITATION = HEP-TH 0401100;%%
}

% Asymmetric Orbifold Refs not present above

\lref\NarainQM{
  K.~S.~Narain, M.~H.~Sarmadi and C.~Vafa, ``Asymmetric Orbifolds,''
  Nucl.\ Phys.\ B {\bf 288}, 551 (1987).
%%CITATION = NUPHA,B288,551;%%
}
\lref\NarainMW{
  K.~S.~Narain, M.~H.~Sarmadi and C.~Vafa, ``Asymmetric Orbifolds:
  Path Integral And Operator Formulations,'' Nucl.\ Phys.\ B {\bf
  356}, 163 (1991).
%%CITATION = NUPHA,B356,163;%%
}
\lref\DineJI{
  M.~Dine and E.~Silverstein, ``New M-theory backgrounds with frozen
  moduli,'' arXiv:hep-th/9712166.
%%CITATION = HEP-TH 9712166;%%
}
\lref\BrunnerFJ{
  I.~Brunner, A.~Rajaraman and M.~Rozali, ``D-branes on asymmetric
  orbifolds,'' Nucl.\ Phys.\ B {\bf 558}, 205 (1999)
  [arXiv:hep-th/9905024].
%%CITATION = HEP-TH 9905024;%%
}
\lref\HWtobe{S.~Hellerman and J.~Walcher, to appear.}

% Algebra and Geometry

\lref\DavidMarks{D.~Marks, work in progress.}

%%%%%%%%%%%%%%%%%%%%%%%%%%%%%%
%%%       Title Page       %%%
%%%%%%%%%%%%%%%%%%%%%%%%%%%%%%

\Title{\vbox{\baselineskip12pt\hbox{hep-th/0602025}
\hbox{MIT-CTP-3718}\hbox{BRX TH-573}\hbox{UPR-1144-T}}}
{\vbox{\centerline{D-Branes in Nongeometric Backgrounds}}}
\centerline{Albion Lawrence${}^1$, Michael B. Schulz${}^2$, and Brian
Wecht${}^3$}\smallskip
\centerline{${}^1$ {\it Brandeis Theory Group, Martin Fisher School of
Physics, Brandeis University}} 
\centerline{{\it MS 057, PO Box 549110, Waltham, MA 02454,
USA}}\smallskip 
\centerline{${}^2$ {\it Department of Physics and Astronomy,
University of Pennsylvania}}
\centerline{{\it Philadelphia, PA 19104, USA}}\smallskip
\centerline{${}^3$ {\it Center for Theoretical Physics, MIT, Cambridge
MA 02139, USA}}\bigskip\bigskip

``T-fold" backgrounds are generically-nongeometric compactifications
of string theory, described by $T^n$ fibrations over a base $N$ with
transition functions in the perturbative T-duality group.  We review
Hull's doubled torus formalism, which geometrizes these backgrounds,
and use the formalism to constrain the D-brane spectrum (to leading
order in $g_s$ and $\alpha'$) on $T^n$ fibrations over $S^1$ with
$O(n,n;\QZ)$ monodromy.  We also discuss the (approximate) moduli
space of such branes and argue that it is always geometric.  For a
D-brane located at a point on the base $N$, the classical
``D-geometry'' is a $T^n$ fibration over a multiple cover of $N$.

\Date{February 2006}

%%%%%%%%%%%%%%%%%%%%%%%%%%%%%%%%%%%
%%%       1. Introduction       %%%
%%%%%%%%%%%%%%%%%%%%%%%%%%%%%%%%%%%

\newsec{Introduction}

In $\CN=1$, $d=4$ compactifications of type IIB with flux,
superpotentials for the complex structure moduli are induced by
turning on various combinations of 3-form flux
\refs{\GukovYA\TaylorII\VafaWI\KachruHE\FreyHF\TripathyQW
\LawrenceZK-\DeWolfeNS}.  In type IIA string theory \refs{\DeWolfeUU,
\GrimmUA}, all geometric moduli can be stabilized by NS-NS and RR fluxes.  In
particular, \hbox{NS-NS} 3-form flux together with nonvanishing
torsion generates a superpotential for the complex structure at tree
level \refs{\LawrenceKJ,\KachruJR}.\foot{This also holds for the
heterotic string, since it involves the common NS-sector.
\refs{\BeckerYV\BeckerGQ-\BeckerSH}.}   In addition, ``geometric''
flux can generate terms in the superpotential coupling K\"ahler and
complex structure moduli \refs{\HullHK\VilladoroCU \DerendingerJN
\CamaraDC-\SheltonCF}.

In the absence of IIA NS-NS flux, the IIA complex structure moduli and
IIB K\"ahler moduli are exchanged by mirror symmetry.  Thus the mirror
of NS-NS flux, when it exists, provides a natural mechanism in type
IIB for stabilizing the K\"ahler moduli
\refs{\LawrenceKJ,\SheltonCF}. The resulting compactifications, and
other non-Calabi-Yau geometries, have been the subject of much recent
interest \refs{\CardosoHD \CardosoAF \SeroneSV \BeckerII \AlexanderEQ
\MicuTZ\ChuangQD-\LindstromIW}.

For Calabi-Yau manifolds with special Lagrangian $T^3$ fibrations,
mirror symmetry corresponds to T-duality along the $T^3$
\refs{\StromingerIT}.  T-duality maps NS-NS 3-form flux $H$ to
geometric ``twisted torus" backgrounds, or to nongeometric backgrounds
\refs{\SheltonCF,\KachruSK\ScherkZR \KaloperYR-\DallAgataFF}.
If $H$ has one index polarized along this $T^3$, the mirror manifold
\refs{\GranaBG,\TomasielloBP} is a ``half-flat" $SU(3)$-structure manifold
\refs{\GurrieriIW \GurrieriWZ \GurrieriDT \deCarlosKH
\FidanzaZI \FreyRN-\HouseYC}. If $H$ has two or three
indices polarized along this $T^3$, the resulting T-dual is not
geometric. In the particular case that two indices of $H$ are
polarized along the $T^3$, the T-dual is known to be a $T^2$ fibration
over $S^1$ with a nongeometric monodromy in the T-duality group
$O(2,2;\QZ)$.  This corresponds to a class of nongeometric flux
studied in \refs{\KachruSK\SheltonCF\GranaNY\BouwknegtVB\BouwknegtWP
\MathaiQQ\MathaiQC\BouwknegtAP-\MathaiFD}.  Such nongeometric string
theory backgrounds are also interesting in their own right and remain
relatively unexplored.\foot{In the case where $H$ has three indices
polarized along the $T^3$, the T-dual is not even locally geometric
\refs{\SheltonCF,\DabholkarVE}.}

D-branes in such backgrounds are central to the physics of these
models. They can provide nontrivial gauge dynamics; space-filling
branes can lead to open string gauge fields, or wrapped branes can
behave as W-bosons.  In addition, D-branes provide an alternate
definition of the background geometry in terms of the moduli space of
the probe.  Wrapped D-brane states are important clues to the
nonperturbative structure of the theory; they constitute part of the
nonperturbative spectrum important for studying string duality, and
they indicate what vacua are connected in field space
\refs{\LawrenceKJ,\GreeneHU}. Finally, D-branes are crucial in our
understanding of mirror symmetry \refs{\StromingerIT,\homologicalmirror}.

In this paper we study D-branes in ``T-fold" (also known as
``monodrofold") compactifications, which are a generalization of the
nongeometric compactifications discussed above.  T-folds are $T^n$
fibrations over a base $N$ with transition functions in the
perturbative T-duality group $G_T$.  While T-folds are locally
describable as geometric manifolds on $N$, there is not a globally
geometric description.  Such manifolds have been studied by a number
of authors \refs{\DabholkarVE\DabholkarSY\HullIN\FlournoyVN\FlournoyXE
\HellermanAX-\LoweQY}.  For bosonic strings, Hull \refs{\HullIN}\ has
proposed a geometrization by doubling the $T^n$ to $T^{2n} \supset
T^n$, which linearizes the $G_T = O(n,n;\QZ)$ action.  The
\hbox{D-branes} on $T^n$ are specified by \hbox{$n$-dimensional}
submanifolds of $T^{2n}$.\foot{This description of D-branes is much
like that for generalized Calabi-Yau backgrounds
\refs{\HitchinUT\GualtieriThesis-\GrangeNM}.}  While the worldsheet
quantum mechanics of \refs{\HullIN}\ is not well-understood, we
already find interesting restrictions on the D-brane spectrum at the
classical level, i.e., to lowest order in $\alpha'$ and $g_s$.

We will focus on T-fold fibrations over $S^1$.  The classically
allowed, locally geometric D-branes for these models are
straightforward to classify.  When the brane wraps the base $S^1$, the
monodromy projects out large classes of D-branes.  When the brane is
at a point on the $S^1$, there is no such projection.  In the latter
case, we find that the classical moduli space of the D-branes is
always a geometric $T^n$ fibration---the Dirichlet $T^n$ bundle---over
a multiple or infinite cover of the $S^1$.  Similarly, for branes
(multiply) wrapping the $S^1$, the classical moduli space is the space
of Wilson lines on the Neumann $T^n$ bundle over (a multiple cover of)
the $S^1$ and is again geometric.  These statements generalize in the
natural way for D-branes in an arbitrary T-fold.

The plan of this paper is as follows.  In \S2\ we review Hull's
formalism, developed in \refs{\HullIN}, for describing twisted torus
compactifications, with particular attention paid to torus fibrations
over a circle.  This review is a partial reworking of Hull's
discussion and makes explicit some points that were implicit in
\refs{\HullIN}.  In \S3\ we review D-branes in these backgrounds,
describe the allowed topological classes of geometric D-branes for
torus fibrations over a circle, and discuss the moduli space of
D0-branes.  We then discuss two examples in particular: $T^3$ with
nonvanishing NS-NS 3-form flux, together with the T-duals described by
\refs{\SheltonCF,\KachruSK,\LoweQY}; and asymmetric orbifolds
describable as T-folds \refs{\DabholkarSY,\FlournoyVN,\FlournoyXE}.
In \S4\ we conclude by presenting some possible directions for further
study.

%%%%%%%%%%%%%%%%%%%%%%%%%%%%%%%%%%%%%%%%%%%%%%%%%
%%%       2. Review of Hull's Formalism       %%%
%%%%%%%%%%%%%%%%%%%%%%%%%%%%%%%%%%%%%%%%%%%%%%%%%

\newsec{Review of Hull's Formalism}

Hull \refs{\HullIN}\ has studied a class of perturbative string
compactifications which can be described as $T^n$ fibrations over a
base $N$.  To construct these fibrations, we begin by compactifying a
D-dimensional string theory down to $D-n$ dimensions on $T^n$.  The
\hbox{T-duality} symmetries of these compactifications are the basis of
T-fold constructions.  In bosonic string theory, with $D=26$, this
compactification has a perturbative T-duality group $G_T =
O(n,n;\QZ)$. The set of large diffeomorphisms $GL(n,\QZ)$ of $T^n$ is
a subgroup of $O(n,n;\QZ)$.  For type II strings ($D=10$), some
elements of the the $O(n,n;\QZ)$ group exchange type IIA and IIB
theories.  These are not symmetries of the worldsheet theory; in these
cases we will take $G_T$ to be the subgroup $SO(n,n;\QZ)$ which maps
IIA(B) to IIA(B).\foot{This group can be identified by studying the
generators of $O(n,n;\QZ)$ as in \S2.4\ of \refs{\GiveonRev}.  These
generators are: large diffeomorphisms of the torus, which have
determinant 1 as elements of $O(n,n;\QZ)$; shifts of the B-field which
have determinant 1; and T-dualities along any leg which have
determinant $-1$.  The elements of $O(n,n;\QZ)$ which take type IIA(B)
back to type IIA(B) must be made up of group elements which have an
even number of T-dualities when written in terms of the generators.
These are all the group elements with determinant $1$.}

In the models we study, we compactify further on a base manifold $N$.
The moduli of the $(D-n)$-dimensional string compactification are
permitted to vary over $N$, so that the full string background
satisfies the worldsheet beta function equations.  One starts by
covering $N$ with a set of coordinate patches $U_{\alpha}$.  For any
two intersecting patches $U_{\alpha},U_{\beta}$ the background fields
on $T^n$ must be identified on $U_{\alpha\beta} = U_{\alpha}\cap
U_{\beta}$, by an action of the structure group.  When the structure
group is the semidirect product of $GL(n,\QZ)$ and $U(1)^n$ (the
latter being the translations in the $T^n$ directions), the resulting
fibration is geometric.  T-folds correspond to fibrations with
structure group $G_T$.  For the remainder of this paper, we will take
$G_T$ to be $O(n,n;\QZ)$ or $SO(n,n;\QZ)$.

Let us first describe the essential point of \refs{\HullIN}, for the
bosonic string theory.  Hull geometrizes the action of $O(n,n;\QZ)$ by
introducing a ``doubled torus" $T^{2n}$.  There is a signature $(n,n)$
metric $L$ on $T^{2n}$, and the ``physical" $T^n$, describing the
target space of the string, is a null submanifold with respect to
$L$. $O(n,n;\QZ)$ is the set of linear transformations on $T^{2n}$
which preserves both the periodicities of the coordinates and the
metric $L$.  The T-fold can be described as a geometric fibration of
$T^{2n}$ over $N$, with transition functions in $O(n,n;\QZ)\subset
GL(2n;\QZ)$.  The string compactification is described by a sigma
model on this $(26+n)$-dimensional space, combined with a self-duality
constraint which projects out $n$ of the coordinates to leave us with
a critical string theory.

At present the quantization of the worldsheet theory is only partially
understood at best.  Furthermore, the discussion here and in
\refs{\HullIN}\ does not include worldsheet fermions. These issues
should be addressed. In this paper, however, we confine our attention
to the classical worldsheet theory, for which specifying the equations
of motion and boundary conditions suffice; furthermore, we restrict
our study to the bosonic degrees of freedom.

\subsec{Basic description of the doubled torus}

\subsubsec{Fields and equations of motion}

The essence of \refs{\HullIN}\ is to present the worldsheet equations
of motion and boundary conditions in a manifestly
$O(n,n;\QZ)$-symmetric manner, building on the earlier work of
Refs.~\refs{\CremmerCT\MaharanaMY-\DuffTF}.  One begins with a
Lagrangian for $2n$ fields, and derives equations of motion and
boundary conditions for the fields.  Next, one imposes a self-duality
constraint on the solutions to the equations of motion, which cuts the
degrees of freedom in half.  This classical analysis will suffice at
lowest order in $\alpha'$ and $g_s$.

We wish to describe a $T^n$ fibration over an $(d-n)$-dimensional base
$N$.  We begin with a set of two-dimensional fields
$\QX^I(\sigma^{\alpha}), Y^A(\sigma^{\alpha})$, with
$(\sigma^0,\sigma^1)$ the worldsheet coordinates, $I = 1,\ldots, 2n$,
and $A = 1,\ldots,d-n$.  The classical Lagrangian for the bosonic
coordinates on the flat worldsheet with metric $\eta={\rm diag}(-1,1)$
can be written as
\eqn\classicallag{
	\CL = - \half \CH_{IJ}(Y) \eta^{\alpha\beta}\p_\alpha\QX^I
	\p_\beta \QX^J  
		- \eta^{\alpha\beta} \CJ_{IA}(Y)\p_{\alpha}\QX^I\p_{\beta}Y^A 
		+ \CL_N(Y),
}
with $\CH$ a positive definite metric and $\QX \in \QR^{2n}/\Gamma
\equiv\ T^{2n}$, where $\Gamma \subset R^{2n}$ is a $2n$-dimensional
lattice.  The quantity $\CJ_{IA}dY^A$ determines the connection for
the $T^{2n}$ fibration over $N$ as discussed in
\HullIN.  We assume a flat worldsheet and therefore ignore the dilaton
coupling. Nevertheless, we will have to specify the dilaton since it
transforms nontrivially under T-duality, as we describe below.

Locally on $N$, given a choice of physical subspace $T^n\subset
T^{2n}$, the $T^{2n}$ metric $\CH$ can be written in terms of the
physical $T^n$ metric $G$ and the NS-NS 2-form potential $B$.  Thus,
$G$ and $B$ are completely geometrized to $\CH$ in the doubled
formalism.  There is no antisymmetric tensor term of the form
$\CB_{IJ}\p_0\QX^I\p_1\QX^J$ in the Lagrangian \classicallag, i.e., no
\hbox{$2n$-dimensional} $\CB$-field analogous to the $B$-field in the
standard sigma model.  Such couplings will arise after an additional
self-duality constraint is solved (cf.~Eq.~(2.3) below).

The equations of motion that follow from varying \classicallag\ are
\eqn\xeom{
	\eta^{\alpha\beta}\p_{\alpha}\left(\CH_{IJ}\p_{\beta}\QX^J
		+ \CJ_{IA}\p_{\beta}Y^A\right) = 0.
}
The Lagrangian and equations of motion are $GL(2n;\QZ)$ invariant.
However, one also imposes the self-duality constraint
\eqn\selfduality{
	\p_{\alpha}\QX^I = L^{IJ}\epsilon_{\alpha}{}^{\beta}
	\left(\CH_{JK}\p_{\beta}\QX^K
		+ \CJ_{JA}\p_{\beta}Y^A\right),
}
which breaks the invariance down to $O(n,n;\QZ)$.  Here
$\epsilon_{\alpha\beta}$ is the two-dimensional antisymmetric tensor.
$L^{IJ}$ is a constant, invertible, symmetric matrix, invariant under
$O(n,n;\QZ)\subset GL(2n;\QZ)$.  Eq.~\selfduality\ cuts the number of
on-shell degrees of freedom in half.  The matrix $LH$ has $n$ positive
and $n$ negative eigenvalues; so, physically, the constraint requires
that half of the $2n$ coordinates be left-moving and half
right-moving.

We impose the constraint \selfduality\ after first varying
\classicallag\ to obtain the equations of motion.\foot{To impose
Eq.~\selfduality\ as a constraint in the path integral, via a Lagrange
multiplier, would imply that the configuration space is the space of
classical solutions.  One would like to have a clean Lagrangian
formulation from which both the equations of motion and the
constraints follow.  At the expense of sacrificing manifest 2d Lorentz
invariance, this could be sought in a slightly modified version of
Tseytlin's formalism \refs{\TseytlinNB,\TseytlinVA}, in which the
self-duality constraint \selfduality\ is an equation of motion.}  The
curl of \selfduality\ gives back the equations of motion \xeom, and
therefore vanishes.  In the presence of boundaries, the allowed
boundary conditions follow from \classicallag\ as well; we will
discuss them in \S3.

A nontrivial fibration over $N$ means that nontrivial elements of
$O(n,n;\QZ)$ relate $\CH$, $\CJ$ and $\QX$ over different coordinate
patches of $N$.  Therefore, we must specify the action of the
structure group $O(n,n;\QZ)$ on these quantities.  The matrix
$g^I{}_J$ acts on lower indices from the right, while the inverse
matrix $(g^{-1})^I{}_J$ acts on upper indices from the left.  Define
$L_{IJ}$ to be the inverse of $L^{IJ}$, such that $L_{IJ}L^{JK} =
\delta_I{}^K$.  A transformation in $O(n,n)$ is defined as one which
preserves $L_{IJ}$:
\eqn\preservation{
	(g^t)_I{}^K L_{KM} g^M{}_J \equiv g^t L g =  L_{IJ}.
}
Then $\CH,\QX,\CJ$ transform as
\eqn\transformation{
\eqalign{
	\CH_{IJ} & \to \CH_{IJ}' = (g^t \CH g)_{IJ},\cr
	\QX^I &\to \QX'^I = (g^{-1})^I{}_J \QX^J,\cr
	\CJ_{IA} &\to \CJ'_{IA} = (g^t)_I{}^K \CJ_{KA}.
}}
We require that $g$ preserve the lattice $\Gamma$, which breaks
$O(n,n)$ down to $O(n,n;\QZ)$.

\subsubsec{Choosing a polarization}

At fixed $Y$, we define a ``physical subspace" of $T^{2n}$ to be a
$T^n\subset T^{2n}$ that is null with respect to the metric $L$.  A
choice of physical subspace is equivalent to a choice of $GL(n)$ in
$O(n,n)$, or alternatively, a choice of coordinate basis for which $L$
has the form
\eqn\newL{
	L = \left(\matrix{0 & \id\cr \id & 0}\right),
}
where ${\bf 1}$ is the $n \times n$ identity matrix.  This change of
basis can be implemented by the $(2n\times 2n)$-dimensional matrix
\eqn\projectorfun{
	\Pi = \left(\matrix{\Pi^i{}_I \cr \tPi_{i I}}\right),
}
where $i = 1\ldots n$.  The lowercase indices are in different
positions in these two projectors to indicate that we have decomposed
a $2n$-dimensional representation of $O(n,n)$ into an $n$-dimensional
representation (the upper indices) of $GL(n)$ and a dual (inverse
transpose) $n$-dimensional representation (the lower indices), i.e.,
$2n
\rightarrow n \oplus n^\prime$. As this corresponds to a basis such
that $L$ has the form \newL, the $\Pi,\tPi$ are themselves null by
construction:
\eqn\nullequation{
	L^{IJ}\Pi^i{}_I\Pi^j{}_J = L^{IJ}\tPi_{iI}\tPi_{jJ} = 0.
}

We can write $X$ in terms of the ``physical" coordinates $X^i =
\Pi^i{}_I\QX^I$ and the ``dual" coordinates $\tilde{X}_i =
\tPi_{iI}\QX^I$.  In this coordinate system, the $O(n,n)$ invariant
metric is
\eqn\newinner{
	ds^2_L = d\QX^I d\QX^J L_{IJ} = 2dX^i d\tX_j.
}
To make contact with the standard sigma model, an additional step is
needed beyond the choice of polarization.  We must restrict $\CH$ to
be an $O(n,n)/\bigl(O(n)\times O(n)\bigr)$ coset metric, that is, to
$\CH = V^tV$, where the vielbein $V$ is a representative of this
coset.  Then, in this basis determined by the polarization, $\CH$ can
be written as
\eqn\newmetric{
	\CH = \left(\matrix{ \CH_{ij} & \CH_i{}^j \cr
				\CH^i{}_j & \CH^{ij}}\right) = \left(
					\matrix{	G_{ij} - B_{ik}G^{kl}B_{lj} & B_{ik}G^{kj}\cr
							-G^{ik}B_{kj} & G^{ij}}\right),
}
where $G$ is the metric and $B$ the NS-NS 2-form potential.  After
solving the self-duality constraints \selfduality, one arrives at the
standard sigma model.

\subsubsec{Some $O(2,2;\QZ)$ transformations}

The utility of this formalism is that it makes the $O(n,n;\QZ)$
duality transformations relatively simple: the coordinates transform
linearly in the $2n$-dimensional defining representation of the group.
To illustrate this, we now review a few basic examples of $O(2,2;\QZ)$
transformations on $T^2$, as also discussed in \HullIN.  We will work
in the basis described by \newL--\newmetric.

\bigskip

\item{$\bullet$} {\bf T-duality on a single cycle}.  T-duality exchanges
a cycle of the torus $T$ with a cycle of the dual torus $\tilde{T}$.
These are generated by
\eqn\tdualshift{
	g_{T,1} = \left( \matrix{0 & 0 & 1 & 0 \cr
					0 & 1 & 0 & 0\cr
					1 & 0 & 0 & 0\cr
					0 & 0 & 0 & 1}\right)
\quad{\rm and}\quad
	g_{T,2} = \left( \matrix{1 & 0 & 0 & 0 \cr
					0 & 0 & 0 & 1\cr
					0 & 0 & 1 & 0\cr
					0 & 1 & 0 & 0}\right).
}
On complex tori, this exchanges the complexified K\"ahler structure
$\rho = \int_{T^2}(B + iJ)$ and the complex structure $\tau$.  Note
that $g_{T,1}$ and $g_{T,2}$ are not elements of the $SO(2,2;\QZ)$
T-duality group of type IIA(B) string theory by themselves; that group
consists of elements with even numbers of these generators.\smallskip
\item{$\bullet$} {\bf Shift of B-field}.  It is easy to check that the
following transformation implements a shift of $\int_{T^2} B$ by $N$
units, which is a symmetry of the closed string theory:
\eqn\bfieldshift{
	g_N = \left(\matrix{ 1 & 0 & 0 & 0\cr
					0 & 1 & 0 & 0\cr
					0 & -N & 1 & 0\cr
					N & 0 & 0 & 1}\right).
}\smallskip
\item{$\bullet$} {\bf Geometric} $GL(n,\QZ)$ {\bf transformations of}
$T^n$.  These are imposed by block-diagonal matrices of the form
\eqn\geomshift{
	g_{geom} = \left( \matrix{A & 0\cr 0 & (A^t)^{-1}}\right), }
where $A$ is a $2\times 2$ invertible matrix with integer entries.
Note that these shifts and the $B$-field shifts generate a group of
block-lower-triangular matrices (with the upper right $2\times 2$
block set to zero).
\medskip

\noindent Hull notes that one can describe T-duality via either an
``active" or a ``passive" transformation.  In an active transformation
the polarization $(\Pi,\tPi)$ remains fixed; the metric data $\CH,\CJ$
transform as in \transformation, while the dilaton transforms as
\eqn\dilatontrans{
	e^{\Phi} \to e^{\Phi'} = \left(\frac{\det G'}{\det
	G}\right)^{1/4} e^{\Phi}. 
}
In a passive transformation, the polarization and the dilaton are
taken to transform (the latter just as in the active transformation),
while the metric data stays fixed.  These are physically equivalent;
we will stick to describing T-duality via active transformations.

\subsubsec{Building T-folds}

The class of ``T-fold'' compactifications under study are locally
$T^n$ fibrations over patches of the base manifold $N$.  Globally, the
backgrounds can be described as follows.  Let $N$ be covered by a set
of open, simply connected neighborhoods $U_{\alpha}$.  In each
neighborhood $U_{\alpha}$, the fibration is ``geometric": the
polarization $(\Pi,\tPi)$ is constant, while the metric $\CH$ can
vary.  If two sets $U_{\alpha}$, $U_{\beta}$, with metrics
$\CH_{\alpha},\CH_{\beta}$ overlap in a set
\ $U_{\alpha\beta} = U_{\alpha}\cap U_{\beta}$, the data are related
by an $O(n,n;\QZ)$ transformation $g_{\alpha\beta}$ acting on
$\CH,\CJ$ as in
\transformation:
\eqn\coordinatetrans{
\eqalign{
	\CH_{\beta} & = g_{\alpha\beta}^t \CH_{\alpha} g_{\alpha\beta},\cr
	\CJ_{\beta} & = g_{\alpha\beta}^t \CJ_{\alpha},
}}
while the dilaton transformation is given by \dilatontrans.  These
transformations must be consistent on triple overlaps
$U_{\alpha\beta\gamma} = U_{\alpha}\cap U_{\beta} \cap U_{\gamma}$:
\eqn\overlap{
	g_{\alpha\beta}g_{\beta\gamma} g_{\gamma\alpha} = 1.
}

As we will see in an explicit example, nontrivial transformations $g$
can lead to NS-NS flux, or the geometric and nongeometric fluxes
discussed in \refs{\SheltonCF,\KachruSK}.  In these examples, we will
call ``geometric" fibrations those for which the transition functions
$g$ are generated by elements of $GL(n;\QZ)$ combined with integral
shifts of the $B$-field. These two operations generate a proper
subgroup of $O(n,n;\QZ)$ which does not contain any T-duality
(inversion) transformations. A nongeometric fibration is simply one
which is not geometric by this definition.

The simplest example of a torus fibration is a fibration over $N =
S^1$.  If the base coordinate is $Y = Y + 2\pi$, we simply demand that
the monodromies live in $\CO(n,n;\QZ)$: $\CH(Y + 2\pi) = g^t \CH(Y)
g$, $\CJ(Y + 2\pi) = g^t \CJ$.  For a more general non-simply
connected base $N$, we impose the same requirements for the
monodromies associated to any noncontractible loop.

Once we have such a fibration, we can generate a set of equivalent
string backgrounds by performing fiberwise T-duality on them: over
each element of the base, we transform the metric and dilaton via
\transformation,\dilatontrans, by some $g_D \in \O(n,n;\QZ)$
which is constant along $N$.  For this transformation to be consistent
in the case of topologically nontrivial $O(n,n;\QZ)$ transformations,
one will also have to conjugate the transition functions
$g_{\alpha\beta}$ or the monodromy matrices $g_{2\pi}$ by $g_D$:
\eqn\conjugatemonodromy{
\eqalign{
	g_{\alpha\beta} & \to g_{\alpha\beta}' = g_D^{-1}
	g_{\alpha\beta} g_D,\cr 
	g_{2\pi} & \to g_{2\pi}' = g_D^{-1} g_{2\pi} g_D.
}}

\subsec{A note on $S^1$ fibrations over a one-dimensional base}

The simplest example to contemplate is an $S^1$ fibration over $S^1$.
For example, we can study a fibration with a monodromy which is equal
to a T-duality:
\eqn\onedtdual{
	g = \left(\matrix{0 & 1\cr 1 & 0}\right).
}	
This is only allowed in bosonic string theory: in type II string theory
this would exchange IIA and IIB, and so is not a symmetry of the string theory.

In particular, we can try to build this fibration via an asymmetric
orbifold: if the base $Y$ has circumference $2\pi R_y$, and the fiber
is at the self-dual radius, the orbifold is realized as a $\QZ_2$
shift taking $Y\to Y + \pi R_y$, $\QX \to g^{-1}\QX$. \foot{Note that
the corresponding asymmetric orbifold constructed in
\refs{\FlournoyXE}\ is not modular invariant, as those authors point
out.  However, the background is locally flat and free of fixed
points; in such cases, Hellerman and Walcher argue that a
modular-invariant asymmetric orbifold can always be constructed
\refs{\HWtobe}.  We would like to thank S. Hellerman for sharing the
results of their work.}

\subsec{$T^3$ with NS-NS flux, and its T-duals}

A nice example which illustrates this formalism is a $T^3$ with $N$
units of NS-NS flux, discussed at length in
\refs{\SheltonCF,\KachruSK,\HullIN}.  We will think of this as a $T^2$
fibration over a base $S^1$.  Note that such a $T^3$ is not in general
a solution to the equations of motion; the beta functions for this
theory do not vanish. As noted in footnote 1 of \refs{\HullIN}, we can
embed this in a string compactification by allowing the moduli of the
$T^3$ to vary over additional spacetime directions.

We can describe the $T^3$ by three periodic coordinates $x,y,z$ with
periods $1$.  We will call $x$ the base coordinate and let $y,z$
correspond to the $T^2$ fibers.  The $T^3$ can be described in the
doubled torus formalism if we write
\eqn\doubledtorus{
	\left(\matrix{ X^i\cr \tX_i}\right) = \left(\matrix{y \cr z \cr \ty \cr \tz}\right)
}
as the doubled fiber coordinates.
 
We will consider constant flux, and choose a gauge such that $B$ is
polarized entirely in the $(y,z)$ directions, so that $B_{yz} = Nx$.
As $x\to x+1$, $B$ shifts by $N$ units.  The monodromy matrix is given
by \bfieldshift.  If the metric $G$ is flat and diagonal, the full
metric $\CH$ can be written as
\eqn\Hforflux{
	\CH = \left( \matrix{ 1 +(Nx)^2 & 0 & 0 & Nx\cr
					0 & 1+(Nx)^2 & -Nx & 0\cr
					0 & -Nx & 1 & 0\cr
					Nx & 0 & 0 & 1} \right).
}

A single T-duality acting on the $y$ direction of this fibration
transforms the monodromy matrix \bfieldshift\ to
\eqn\oncetdual{
	g_{NT} = g_T^{-1} g_N g_T =
	\left( \matrix{1 & -N & 0 & 0\cr
			0 & 1 & 0 & 0\cr
			0 & 0 & 1 & 0\cr
			0 & 0 & N & 1}\right),
}
and the corresponding T-dual metric is
\eqn\HforfluxT{
	\CH_T = g_T^t \CH g_T = \left(\matrix{1 & -N & 0 & 0\cr
								-N & 1 + (Nx)^2 & 0 & 0\cr
								0 & 0 & 1+(Nx)^2 & N\cr
								0 & 0 & N & 1}\right),
}
corresponding to a vanishing $B$-field and a metric of the form
\eqn\onetmetric{
	ds^2 = dx^2 + (dy - N x dz)^2 + dz^2.
}
This is the twisted torus background supporting ``geometric" flux as
described in \refs{\SheltonCF,\KachruSK}.

In the doubled torus picture, the {\it components\/} of $\CH$
transform by the monodromy element \oncetdual, however, the {\it
tensor\/} $\CH$ is single-valued on the $T^{2n} = T^4$ fibration.
Since the example discussed in this section is geometric, the
monodromy satisfies the more restrictive properties that it acts only
on the components of $G$ and that the tensor $G$ is single-valued on the
geometrical $T^n = T^2$ fibration.  From this single-valuedness of
$G$, we deduce the coordinate transition functions discussed in
\refs{\KachruSK}:
\eqn\newperiodicity{
	(x,y,z) \sim (x,y+1,z)\sim(x,y,z+1)\sim (x+1,y+Nz,z).
}

We can now imagine performing a second T-duality, this time along the
$z$ direction.  It is not immediately clear from the above discussion
that this is possible: the isometry along the $z$ direction seems to
be broken by \newperiodicity.  On the other hand, if we started with
the original $T^2$ fibration, for which the periodicities and the
metric were invariant under infinitesimal shifts in the $(y,z)$
direction, there would seem to be nothing stopping us from performing
a duality transformation
\eqn\doublet{
	g_{T2} = \left(\matrix{ 0&0&1&0\cr
					0&0&0&1\cr
					1&0&0&0\cr
					0&1&0&0}\right).
}
If we perform a T-duality by this matrix, fiber by fiber, the
monodromy matrix becomes
\eqn\newmonod{
	g_{NT2} = g_{T2}^{-1} g_N g_{T2} =
	\left(\matrix{1&0&0&-N\cr
			0&1&N&0\cr
			0&0&1&0\cr
			0&0&0&1}\right).
}
This is a nontrivial element of $O(2,2;\QZ)$.
The metric $\CH$ becomes
\eqn\Htwotduals{
	\CH_{NT2} = g_{T2}^t \CH g_{T2} =
	\left(\matrix{ 1&0&0&-Nx\cr
			0&1&Nx&0\cr
			0&Nx&1+(Nx)^2&0\cr
			-Nx&0&0&1+(Nx)^2}\right),
}
corresponding to a NS-NS $B$ field of the form $B_{yz} =
Nx/(1+(Nx)^2)$ and a metric of the form
\eqn\twotdualmetric{
	ds^2 = dx^2 + \frac{dy^2 + dz^2}{1 + (Nx)^2}.
}
In other words, the volume of the torus shrinks as one moves around
the $S^1$.  This is a candidate for a nongeometric compactification.
While it is locally geometric, the fibration globally has a monodromy
which is a nontrivial element of the T-duality group.

%%%%%%%%%%%%%%%%%%%%%%%%%%%%%%%%%%%%%%%%%%
%%%       3. D-branes on T-folds       %%%
%%%%%%%%%%%%%%%%%%%%%%%%%%%%%%%%%%%%%%%%%%

\newsec{D-branes on T-folds}

\subsec{Geometric D-branes on $T^n$}

Consider a worldsheet $\Sigma$ with boundary $\p\Sigma$.  We will
assume that the boundary conditions are those derived by varying the
action \classicallag, and demand that they be consistent with the
self-duality constraints \selfduality.

We parameterize the boundary by the worldsheet coordinate $\sigma^0$,
and the normal direction by $\sigma^1$.  The boundary terms that arise
from varying the action \classicallag\ with respect to $\QX$ are
\eqn\boundaryterm{
	\delta S_{\rm bound} = \delta\QX^I \left( \CH_{IJ} \p_1 \QX^J
	+ \CJ_{IA}\p_1 Y^A\right)|_{\p\Sigma} 
	= 0.
}
Let $\Pi^I_{D,J}$ be a projector onto Dirichlet directions, i.e.,
directions perpendicular to the Dirichlet branes.  Then, $\Pi_D
\delta \QX = \delta \QX \Pi_D^t = 0$.  Using $\id = \Pi_D^t + (1 -
\Pi_D^t)$, and defining $(\id - \Pi^t_D)_J{}^I\equiv \Pi_{N,J}{}^I$,
we find that
\eqn\betterboundary{
	\delta S_{\rm bound} = \delta \QX^I \Pi_{N,I}{}^J \left(
		\CH_{JK}\p_1\QX^K 
		+ \CJ_{JA} \p_1 Y^A\right) = 0.
}
or 
\eqn\neumanncondi{
	\Pi_{N,I}{}^J \left( \CH_{JK}\p_1\QX^K + \CJ_{JA} \p_1 Y^A\right) = 0.
}
Note that the Dirichlet and Neumann directions are orthogonal with
respect to $\CH$.  Note also that the D-brane is not completely
specified by $\Pi_D$: only the directions perpendicular to the brane
are.  The actual position of the brane is undetermined at this level.

These boundary conditions are consistent with the self-duality
constraint \selfduality\ if
\eqn\sdboundarycond{
	\Pi_{N,I}{}^J L_{JK} \p_0\QX^K = 0,
}
which is guaranteed if
\eqn\intertwining{
	\Pi_{N,I}{}^J L_{JK} = L_{IJ}\Pi^J_{D,K}.
}

Eq.~\intertwining\ implies (as asserted in \refs{\HullIN}) that the
D-branes are null with respect to $L$: Using the fact that
\eqn\piannn{
	\Pi_D^t \Pi_N = \Pi_D^t (\id - \Pi_D^t) = 0,
}
we can multiply \intertwining\ on the left by $\Pi_D^t$ or on the
right by $\Pi_N^t$ to find the conditions
\eqn\nullbraneconditions{
\eqalign{
	\Pi_D^t L \Pi_D & = 0,\cr
	\Pi_N L \Pi_N^t & = 0.
}}
These two conditions are not identical (as we will see in an example)
and both must be imposed.  Additionally, the boundary conditions imply
that the classical stress tensor is conserved at the boundary:
\eqn\Tzeroone{T_{01}|_{\p\Sigma} = 0.}
This remains true for the self-dual subset of solutions to the
equations of motion.

Eq.~\sdboundarycond\ indicates that for each Neumann condition, there
is a Dirichlet condition related by an action of $L$.  Therefore,
there are always $n$ Neumann directions and $n$ Dirichlet directions
on the doubled torus $T^{2n}$---the rank of $\Pi_N$ and $\Pi_D$ must
each be $n$---and these directions each form a null $T^n$ subspace of
$T^{2n}$, in the sense of \nullbraneconditions.  Globally, $\Pi_N$ and
$\Pi_D$ each define a projection from the $T^{2n}$ bundle to a null
$T^n$ bundle over $N$.  However, there is an important distinction
between the Neumann and Dirchlet $T^n$ bundles.  Only the Neumann
$T^n$ bundle, which is wrapped by the brane, need exist as a subbundle
of the $T^{2n}$ bundle.
\smallskip

In summary: A D-brane in the doubled torus formalism wraps a null
$T^n\subset T^{2n}$ subbundle over a cycle $S$ of the base $N$.
\smallskip
Note that this entire discussion is at lowest order in $\alpha'$ and
$g_s$, which is the order at which we understand the doubled torus
formalism.  There is no guarantee these branes will be stable once
higher-order corrections are included.

\subsubsec{T-duality acting on D-branes}

We have seen that D-branes are specified by the Dirichlet projector
$\Pi^I_{D,J}$.  The index structure determines the transformation
properties of this brane under an $O(n,n;\QZ)$ transformation $g$:
\eqn\dualityonbranes{
	\Pi_D \to \Pi_D' = g^{-1} \Pi_D g.
}

\subsubsec{Example: Boundary states in the $c=1$ Gaussian model and
``nongeometric" branes} 

The $c=1$ Gaussian model corresponding to a target space circle at
radius $R$ is described in our formalism as a $T^2$ with metric
\eqn\conemetric{
	\CH = \left( \matrix{ R^2 & 0 \cr 0 & R^{-2} }\right).
}
The null directions are specified by a matrix $\Pi_D$ which: (i) is a
projector, $(\Pi_D)^2 = \Pi_D$; and (ii) satisfies
\nullbraneconditions.  A simple calculation reveals that the solutions
to the first equation in \nullbraneconditions\ are:
\eqn\coneprojectors{
\eqalign{
	\Pi^{(1)}_D(a) & = \left(\matrix{1 & a \cr 0 & 0}\right),\cr
	\Pi^{(2)}_D(b) & = \left( \matrix{0 & 0 \cr b & 1}\right).
}}
These do not specify physical boundary conditions after \selfduality\
is solved for $X$.  For example, $\Pi^{(1)}(a)$ leads to the boundary
conditions
\eqn\firstbctype{
\eqalign{
	& \phantom{-} \delta X + a \delta \tX = 0,\cr
	& - a R^2 \p_1 X + \frac{1}{R^2} \p_1 \tX = 0 \Rightarrow
		- a R^{-2} \p_1 X + \p_0 X = 0.
}}
However, from the second equation in \nullbraneconditions, we find
that only $\Pi^{(1)}_D(0),\Pi^{(2)}_D(0)$ are allowed.  These
correspond to fully Dirichlet or fully Neumann conditions on the
circle. Note that $\Pi_D$ does not fully specify the moduli of these
D-branes; the D0-brane can be at any point on the circle, and the
Wilson line on the D1-brane can have any value.  T-duality in this
case is described by the matrix
\eqn\onedtdualitymatrix{
	g = \left( \matrix{ 0 & 1\cr 1 & 0}\right).
}
It is easy to see that this transforms Dirichlet into Neuman
conditions, and vice-versa.

Friedan \refs{\Friedannotes} has characterized the space of conformal
boundary conditions for $c=1$ conformal field theories.  These
boundary states have been constructed in
\refs{\GaberdielZQ\JanikHB\CappelliWQ-\TsengAX}.  They consist of the
Dirichlet and Neumann conditions, together with a continuous family of
additional boundary states, which have finite boundary entropy (and
thus finite tension) only when the radius of the circle is a rational
number times the self-dual radius \refs{\TsengAX}.  These additional
boundary states appear to be deformations of single and multiple
D-branes by boundary potentials.  By moving in the family of boundary
states we may interpolate between Dirichlet and Neumann conditions.
It would be interesting to describe these states in the doubled torus
formalism; in the open string channel, they would require including
explicit boundary terms in the Lagrangian of the theory.

\subsec{Torus fibrations}

We would now like to study D-branes on $T^n$ fibrations over an $S^1$
base, inspired by the example of a $T^3$ with NS-NS 3-form flux.  In
this case there are restrictions on the allowed D-branes, due to the
nontrivial monodromy.  As we have seen above, geometric D-branes on a
doubled torus $T^{2n}$ correspond to null $n$-planes.  If these are to
have finite energy, they must wrap an $n$-cycle of the $T^n$.  These
cycles are topologically distinct and are labeled by integers.

If the D-brane sits at a point on the base $S^1$ there are no
restrictions on this $n$-plane, other than that it corresponds to a
physical, conformally-invariant boundary condition.  As noted by Hull
\refs{\HullIN}, if we transport this brane a full period about the
$S^1$, the metric has changed but the topological type of the D-brane
has not (the open string moduli---the position of the D-brane, or the
value of a Wilson line modulus---may, of course, change).  A monodromy
transformation will act on $\CH,\CJ$ as in \transformation.  It will
also act on $\Pi_D$ as in \dualityonbranes. Therefore, D$p$-branes can
transmute into D$p'$-branes as they go once around the base, or their
dimensions may remain constant, but the topological class of the
$T^{n}$ which they wrap will not change.  For such D-branes, the
classical moduli space\foot{By ``classical moduli space'' we mean the
space of classically allowed boundary conditions.  At higher order in
$\alpha'$ or $g_s$, the desired probes may break supersymmetry, be
unstable, or develop a potential along the $S^1$, meaning that there
is strictly speaking no moduli space.  Furthermore, in the example
with $H$-flux that we discuss in \S3.4, the full string theory will
have varying dilaton and the probe will experience a potential in the
directions transverse to the torus; the term ``moduli space" is then
used loosely here to refer to approximately flat directions in the
full $\alpha'$ and $g_s$ corrected theory.  In the weak-coupling
region of such configurations, the space of classical boundary
conditions should be one good measure of the geometry seen by the
D-brane probe.} of positions on the base is not the $S^1$ but rather
an $m$-fold cover,\foot{Here we are ignoring the $\QZ_2$ issue of
orientation (cf.~\S3.5).  In some cases orientation considerations
require a moduli space that is a $2m$- rather than $m$-fold cover of
$S^1$.} where $m$ is the smallest integer such that $g^m\Pi_D g^{-m}
= \tPi_D$ imposes the same boundary conditions as $\Pi_D$.  (As we
will see in an example below, this does not require that $\Pi_D =
\tPi_D$. Rather, it merely requires that the row vectors of the matrix
$\tPi_D$ be linear combinations of the row vectors of $\Pi_D$.)  It is
possible that $m$ is infinity.  When one includes the moduli from
Wilson lines and transverse positions on the physical $T^n$ fiber, the
classical moduli space is conveniently described in a
polarization-independent manner as the Dirichlet $T^n$ fibration over
the cycle $mS^1$; likewise, for D-branes located at points on the base
$N$ in a more general T-fold, the space of transverse positions on the
base is $mN$ for some $m$, and the classical D-brane moduli space is
the Dirichlet $T^n$ fibration over $mN$.\foot{This statement holds
only for the classical moduli space, in the sense of Footnote~7.
Since the true moduli space of Wilson lines is the space of {\it flat}
connections, the Wilson lines are lifted for any $S^1$ factors of the
$T^n$ fiber that get a nontrivial $S^1$ connection over $N$.  For
example, in the background \twotdualmetric, we cannot turn on a Wilson
line in the $y$-direction without energy cost, since even for constant
$A$, the field strength $F = d\bigl(A_y (dy-Nxdz)\bigr)$ is
nonzero. This lifting is a 1-loop effect on the worldsheet.}

If the D-brane wraps the base, then a similar argument restricts the
allowed D-branes: The D-brane must close on itself. This cannot happen
if, after one follows the brane around the circle, the boundary
conditions are not invariant under the monodromy transformation.
However, it is possible that the D-brane is wrapped $m$ times around
the base, e.g. it only closes in on itself after one goes around the
circle $m$ times.  In this case, we need simply demand that there
exists an $m$ where
\eqn\allowedbasewrappings{
	g^{-m} \Pi_D g^m = \tilde \Pi_D.
}
has the same space of zero eigenvectors as the original
projector. Note that for higher-dimensional base manifolds $N$ this
condition holds for any cycle $\gamma$ of $N$, topologically trivial
or nontrivial, which lies inside a D-brane.

For a D-brane that wraps the entire base manifold $m$ times, the
worldvolume in the doubled geometry is Neumann $T^n$ bundle over $mN$,
i.e., the pullback of Neumann $T^n$ bundle under the map $mN\to N$.
The classical moduli space is the space of Wilson lines on this
bundle.  In the case that the base is $S^1$, it is tempting to try to
give this space a more explicit description along the lines of ``the
Dirchlet $T^n$ bundle over the circle dual to $mS^1$.''  However,
within the T-fold formalism, where the fiber alone is doubled, the
latter is not defined.  Rather, the formalism of Ref.~\DabholkarVE,
where both the base and fiber are doubled, is needed to make the
statement more precise.

There is one subtlety with condition \allowedbasewrappings: the
boundary conditions describing the coordinates of the brane may be
invariant under the monodromy transformation, while the orientation of
the submanifold may not be. In the case that the fibration is
geometric, the corresponding D-brane would be wrapping an unoriented
surface.  This is an issue for type II D-branes: such D-branes will
only be allowed in the presence of an appropriate orientifold.  Other
than a couple of comments in \S3.5, we will leave such additional
consistency conditions for future work.

\subsec{The nonperturbative topology of T-folds}

%\ShenkerXQ
\lref\ShenkerXQ{
  S.~H.~Shenker, ``Another Length Scale in String Theory?,''
  arXiv:hep-th/9509132.
%%CITATION = HEP-TH 9509132;%%
}
%\KabatCU
\lref\KabatCU{
  D.~Kabat and P.~Pouliot, ``A Comment on Zero-brane Quantum
  Mechanics,'' Phys.\ Rev.\ Lett.\ {\bf 77}, 1004 (1996)
  [arXiv:hep-th/9603127].
%%CITATION = HEP-TH 9603127;%%
}
%\DouglasYP
\lref\DouglasYP{
  M.~R.~Douglas, D.~Kabat, P.~Pouliot and S.~H.~Shenker, ``D-branes
  and short distances in string theory,'' Nucl.\ Phys.\ B {\bf 485},
  85 (1997) [arXiv:hep-th/9608024].
%%CITATION = HEP-TH 9608024;%%
}

At weak string coupling, D-branes are known to be excellent
nonperturbative probes of distances below the string scale
\refs{\ShenkerXQ\KabatCU-\DouglasYP}.  Therefore, they would seem to
be an excellent probe of ``stringy" nongeometric compactifications.
``D-geometry", as opposed to ``stringy" geometry, typically refers to
the moduli space of the D-brane probe.  It could also refer to the
configuration space of the worldvolume theory of the probe, as
accessed by finite-energy scattering \refs{\DouglasYP}.

What we find\foot{As do the authors of \refs{\DouglasYP}\ in studying
finite-energy D0-brane scattering in ALE spaces.}\ is that there is
also a ``D-topology" which is distinct from the ``stringy" topology of
the sigma model target space.  In particular, the stringy topology for
$T^n$ fibrations over $S^1$ is completely defined by the $O(n,n;\QZ)$
monodromy, and is nongeometric if $g$ is not generated by $GL(n;\QZ)$
transformations and $B$-field shifts.  However, the classical moduli
space of D0-branes will always be described by fibrations with
geometric monodromies.

To see this, imagine a D0-brane on a $T^n$ fibration
of $S^1$ with a nongeometric monodromy $g$.  Transport the D0-brane
around the $S^1$ base; it will return to its image as defined by
$\tPi_D = g^{-1}\Pi_D g$. Such a monodromy will transform a D0-brane
into some other brane. The moduli space is at best a multiple cover of
the fibration.  If there is a smallest $m$ such that $g^{-m} \Pi_D
g^m$ has the same invariant subspace as $\Pi_D$, then the D0-brane
will return to itself after being transported around the circle $m$
times.  However, in this case $g^m$ cannot be a T-duality monodromy,
or it would not leave the D0-brane invariant.  It must therefore be a
geometric monodromy, a combination of $GL(n;\QZ)$ transformations of
the $T^n$ and $B$-field shifts.  If the $S^1$ has radius $R$, the
moduli space of the D0-brane will be a geometric $T^n$ fibration---the
Dirichlet $T^n$ bundle---over a circle of radius $mR$.

\subsec{Example: $T^3$ with $H$-flux and its T-duals}

For our first example we will return to the $T^3$ with flux studied in
\S2.3.  The monodromy matrix is given in \bfieldshift, and we can use
this to check whether the invariant subspace of particular Dirichlet
projectors survives the monodromy transformation.

First, pick both Dirichlet directions in the ``physical" $X^i$
directions, corresponding to
\eqn\dirone{
\Pi_{D,1} = \left (
\matrix{ \id & {\bf 0} \cr {\bf 0} & {\bf 0}} 
\right ),
}
where $\id$ is the $2\times 2$ identity matrix and ${\bf 0}$ is a
$2\times 2$ block of zeroes. $\Pi_D$ specifies a brane sitting at a
point in the fiber; if the brane wraps the base, it is a
$D1$-brane. Upon encircling the base, $\Pi_{D,1}$ transforms as
\eqn\dironetrans{
\widetilde \Pi_{D,1} = g^{-1} \Pi_{D,1} g = \left( 
\matrix{1 & 0 & 0 & 0 \cr
0 & 1 & 0 & 0 \cr
0 & N & 0 & 0 \cr
-N & 0 & 0 & 0} \right).
}
Since the third and fourth rows of $\tPi_D$ are simply multiples of
the second and first rows respectively, the two projectors have the
same invariant subspace and define equivalent D-branes.  The original
boundary conditions are $\Pi_{D,1} \partial_0 \QX^I = \partial_0 X^i =
0$, and the transformed boundary conditions are
\eqn\dironebc{
\widetilde \Pi_{D,1} \partial_0 \QX^I = \left ( \matrix{\partial_0 y
\cr \partial_0 z \cr N\partial_0 z \cr -N\partial_0 y} \right ) =0,
}
which are equivalent.  Thus we see that a D1-brane wrapping the base
is allowed.
 
Next, consider a D2-brane wrapping the base and the $y$ direction in
the fiber, and take the Dirichlet projector to be along the $y$ and
$\widetilde z$ directions,
\eqn\dirtwo{
\Pi_{D,2} = \left(
\matrix{ 1 & 0 & 0 & 0 \cr 
0 & 0  & 0 & 0 \cr
0 & 0 & 0 & 0 \cr
0 & 0 & 0 & 1} 
\right).
}
It is easy to check that in this case, $\widetilde \Pi_{D,2} =
\Pi_{D,2}$. This D2-brane is allowed, since the boundary conditions
are trivially the same. Similarly, the projection onto the $z$ and
$\widetilde y$ directions gives an allowed D2-brane. One can also
check, however, that the projection onto the $y, \widetilde y$ or $z,
\widetilde z$ directions does not give invariant boundary conditions.
 
Finally, consider the case of a D3-brane wrapping both directions of
 the fiber. The projector here is
\eqn\dirthree{
\Pi_{D,3} = \left(
\matrix{ {\bf0} & {\bf0} \cr {\bf0} & {\bf1}} 
\right),
}
which transforms to
\eqn\dironetrans{
\widetilde \Pi_{D,3} = g^{-1} \Pi_{D,3} g = \left( 
\matrix{0 & 0 & 0 & 0 \cr
0 & 0 & 0 & 0 \cr
0 & N & 1 & 0 \cr
-N & 0 & 0 & 1} \right).
}
The original boundary conditions were $\Pi_{D,3} \partial_0 \QX^I =
\partial_0 \widetilde X^i = 0$, but the transformed boundary
conditions are $ -N \partial_0 z+ \partial_0 \widetilde y = N
\partial_0 y + \partial_0 \widetilde z =0$. These are no longer
satisfied, so we see that a D3-brane wrapping the entire $T^3$ is not
allowed.

It is important to realize that the correct definition of the Neumann
boundary conditions is {\bf not} $\Pi_N \partial_1 \QX = 0$, but
instead $\Pi_N \CH \partial_1 \QX = 0$, as implied by varying
\classicallag. For example, let us study one of the above D-branes
wrapping the $S^1$.  The Dirichlet conditions $\Pi_D \p_0 \QX = 0$ are
invariant under the monodromy, as are the conditions $\Pi_N \CH \p_1
\QX = 0$.  Since $\CH$ can, and in this case does, depend on the
coordinate along the base, $\Pi_N \p_1 \QX$ will change as we circle
the $S^1$. For the Neumann projector
\eqn\neuone{
\Pi_{N,1} = 1 - \Pi_{D,1}^t = \left(
\matrix{ {\bf 0} & {\bf 0} \cr {\bf 0} & {\bf 1}} 
\right)
}
with $\CH$ specified in \Hforflux, the boundary conditions $\Pi_N
\p_1 \QX$ are $Nx \partial_1 z - \partial_1 \widetilde y = Nx \partial_1 y
+ \partial_1 \widetilde z = 0$.  As $x \rightarrow x + 1$, these
equations will vary and manifestly not have the same solutions.
 
We conclude this example by investigating the backgrounds T-dual to
this one, and checking that the allowed branes are what we would have
na\"{\i}vely expected from T-duality.

\subsubsec{\it One T-duality.}

T-dualizing along the $y$-direction gives a background with monodromy
\oncetdual. Everything is as expected, as can be seen by switching $y
\leftrightarrow \widetilde y$ in the Dirichlet projectors above. The
projections onto the $(y,z)$, $(\widetilde y, z)$, and $(\widetilde
y,\widetilde z)$ give consistent boundary conditions.  The projection
onto the $(y, \widetilde z)$ directions do not.  Thus, we are allowed
to have D-branes of all dimensions in this background.
 
Since this background is geometric---the twisted $T^3$ of
Eq.~\onetmetric---the spectrum of \hbox{D-branes} can be checked
directly.  The cohomology of the twisted $T^3$ is like that of $T^3$
except that the global 1-form in the $y$-direction is $e^y = dy-Nxdz$,
with $de^y = -Ndx\wedge dz$.  So, $dx\wedge dz$ generates a $\QZ_N$
torsion factor in $H^2$ and $e^z\not\in H^1$.  Consequently, the
homology is like that of $T^3$ except that $S^1_y$ is a $\QZ_N$
torsion class, and there is no $xz$-cycle (no section of the $S^1_y$
fibration).  Thus, the possible branes wrapping the $S^1_z$ base are
D1$_x$, D2$_{xy}$ and D3$_{xyz}$, with D2$_{xz}$ not allowed.

In the doubled formalism, these results translate into the geometrical
statement that there exist null subbundles $T^2_{\tilde y\tilde z}$,
$T^2_{y\tilde z}$ and $T^2_{yz}$, but not $T^2_{z\tilde y}$.  This can
be checked directly as well.  The metric of the $T^4$ fibration over
$S^1$ is
\eqn\adblmetric{ds^2 = dx^2 + \bigl((e^y)^2 + dz^2\bigr) +
\bigl({d\tilde y}^2 + (e^{\tilde z})^2\bigr),}
where $e^{\tilde z} = d\tilde z + Nx d\tilde y$.  The quantity in the
first set of parentheses is the physical $T^2$ fiber metric, and that
in the second set of parentheses is the dual (inverse) $T^2$ metric.
Thus, $H^2$ of the doubled geometry contains $e^y\wedge dz$, $dz\wedge
d\tilde y$ and $d\tilde y\wedge de^{\tilde z}$, but not $e^y\wedge
e^{\tilde z}$, since latter is not closed.  The homology dual of this
statement exactly reproduces the correct list of null $T^2$
subbundles.

This example illustrates the fact that the Dirichlet $T^n$ bundle need
not be a subbundle of the $T^{2n}$ bundle.  Consider the D-brane that
wraps the $T^2_{y\tilde z}$ subbundle.  The metric $\adblmetric$
projects to the product metric on $S^1_x\times T^2_{y\tilde z}$ for
the Neumann bundle, and to the product metric on $S^1_x\times
T^2_{z\tilde y}$ for the Dirichlet bundle.  However, only the Neumann
bundle is a subbundle of the doubled geometry.

\subsubsec{Two T-dualities}

Now consider the background with monodromy \newmonod, which we get by
additionally T-dualizing in the $z$-direction. It is straightforward
to check that the allowed projections are onto the $(y,\widetilde z),
(\widetilde y,z)$, and $(\widetilde y,\widetilde z)$ directions,
corresponding to two D2-branes and one D3-brane. Since there were no
D3-branes in the original $H$-flux background, there are no D1-branes
here.\foot{Since it is not a T-fold, the background obtained from the
$T^3$ with $H$-flux after three T-dualities is not discussed in this
paper.  However, note that the absence of D3 branes in the original
background implies that D0-branes do not exist after three
T-dualities.  Therefore, such backgrounds are not even locally
geometric: there are not even points on which to place D0-branes.}

The moduli space of D0-brane positions on the $S^1$ base in this
background is an infinite cover of the $S^1$.  To see this, consider
the image of the D0-brane under T-duality, in the original background
with $H$-flux.  This is a D2-brane wrapping the $T^2$ fiber.  After
transport around the $S^1$ base, the $B$-field will induce $N$ units
of D0-brane charge.

\subsec{Example: asymmetric orbifolds of $T^3$}
 
As a final example, let us consider an orbifold of $T^3$ which
combines a shift on the base with an action on the moduli of the $T^2$
fiber. To begin, consider orbifolding by the action
 
\eqn\tauorb{
\left \{ \matrix{\tau \rightarrow -1/\tau,\cr
x \rightarrow x + 1,} \right. 
}
which implements a $90^\circ$ rotation on the fiber upon encircling
the base. This is a perfectly good geometrical (and symmetric)
orbifold, but will help us with related nongeometric orbifolds
momentarily. It is clear what the allowed D-branes wrapping the base
are: Since this action rotates the two 1-cycles of the fiber, one
expects that a D1-brane wrapping the base or a D3-brane wrapping the
entire fiber will be invariant, but not a D2-brane wrapping the base
once while wrapping a one 1-cycle of the fiber. We can easily see this
from the projectors $\Pi_{D,N}$. The monodromy is simply
\eqn\taumon{
g_\tau = \left (
\matrix{ {\bf S} & {\bf 0} \cr
{\bf 0} & {\bf S}}
\right ),
}
where ${\bf S}$ is a $2\times2$ antisymmetric matrix with 1 in the
upper right corner. It is now an easy matter to show that the
projectors onto both the $(y,z)$ and $(\ty, \tz)$ directions are
invariant under this monodromy, but the projectors onto the $(y,\tz)$
and $(\ty,z)$ directions are not. Thus, D1-branes and D3-branes
wrapping the base are allowed, but D2-branes wrapping the base once
are not.

Since the orbifold \tauorb\ is a geometrical background, we can check
this result directly.  The 3-dimensional geometry is $\QR^3/\Gamma$,
where $\Gamma$ is generated by the group elements $\alpha$, $\beta$
and $\gamma$, taking $(x,y,z)$ to $(x+1,z,-y)$, $(x,y+1,z)$ and
$(x,y,z+1)$, respectively.  To determine which cycles can be wrapped,
we need the homology of $\QR^3/\Gamma$.  This is easily
computed\foot{Clearly, $H_0 = H_3 = \QZ$.  The fundamental group
$\pi_1$ is $\Gamma$, with nonzero commutators
$\alpha\beta\alpha^{-1}\beta^{-1} = \gamma^{-1}\beta^{-1}$ and
$\alpha\gamma\alpha^{-1}\gamma^{-1} = \beta\gamma^{-1}$.  Given
$\pi_1$, the group $H_1(\QZ)$ is obtained by setting all commutators
to unity.  This gives $H_1 = \QZ\oplus\QZ_2$, from $\alpha$ and
$\beta$, respectively.  Finally, $H_2 = \QZ$ by Poincar\'e duality
together with the the Universal Coefficient Theorem $H^{\rm torsion}_n
= H^{n+1}_{\rm torsion}$.}\ to be $H_0 = H_2 = H_3 = \QZ$ and $H_1 =
\QZ\oplus\QZ_2$.  The corresponding $\QZ$-valued classes are a point, the
$T^2$ fiber, the whole manifold, and the $S^1$ base.  The $\QZ_2$
torsion cycle is the class of the $z$ circle fiber, which via the
identification is the same as a circle fiber oriented in the $-z$ or
$\pm y$ directions.

Qualitatively, the geometry can be thought of as an oriented, higher
dimensional, $\QZ_4$ analog of a Klein bottle.  A Klein bottle is an
$S^1$ fibration over $S^1$ with a $\QZ_2$ orientation reversal
twisting the fiber.  Here, we instead have a $T^2$ fibration over
$S^1$ with a $\QZ_4$ rotation twisting the fiber.  In fact, there
exists a Klein bottle within this geometry.  Since ${\bf S}^2 = -1$,
we have $g^{-2}_\tau\Pi_D g^2_\tau = \Pi_D$, so it appears that by
Eq.~\allowedbasewrappings\ a D2 brane can twice wrap the base {\it
twice} while wrapping a 1-cycle of the fiber.  However, this D-brane
is wrapping an unoriented cycle---more precisely, a Klein bottle: Such
a configuration will be allowed in the bosonic string, and certain
orientifolds of the type II string.

T-dualizing this example on one of the fiber directions interchanges
the complex structure and K\"{a}hler moduli.  The result is a
manifestly nongeometric orbifold with action
\eqn\rhoorb{
\left \{ \matrix{\rho \rightarrow -1/\rho,\cr
x \rightarrow x + 1,} \right.
}
which mixes $B$-field and metric.  It is also clearly an asymmetric
orbifold \refs{\DabholkarSY, \FlournoyXE,
\AokiSM\NarainQM\DineJI\BrunnerFJ-\NarainMW}: the combined action
$\tau \to -1/\tau, \rho \to -1/\rho$ is the same as two T-dualities,
so the action $\rho \to -1/\rho$ is just the asymmetric T-duality
action combined with a symmetric $90^\circ$ rotation, and as such is
still asymmetric. Since this orbifold is just one T-duality away from
the $\tau\to -1/\tau$ orbifold where the only allowed branes wrapping
the base once were D1-branes and D3-branes, here we only expect
D2-branes to be allowed. The appropriate monodromy matrix is
\eqn\rhomon{
g_\rho = \left (
\matrix{ {\bf 0} & {\bf S} \cr
{\bf S} & {\bf 0} }
\right ).
}
One can easily check that only the Dirichlet projectors onto the $(y,
\tz)$ and $(\ty, z)$ directions give invariant boundary conditions,
confirming our intuition.

Finally, let us combine these two actions into an asymmetric orbifold
that is not dual to a symmetric orbifold,\foot{In \FlournoyXE, the
na\"{\i}ve presentation of this orbifold for the heterotic string was
shown to not be modular invariant. Whether there exists a consistent
prescription is the subject of current study \HWtobe. We proceed with
this example for illustrative purposes. }
\eqn\rhotau{
\left \{ \matrix{\rho \rightarrow -1/\rho,\cr
\tau \to -1/\tau,\cr
x \rightarrow x + 1.} \right.
}

Consider D-branes that wrap the base exactly once. As stated above,
the action \rhotau\ is the same as completely T-dualizing the
fiber. Thus, we expect no such D-branes to exist in this
background. The appropriate monodromy here is
\eqn\rhomon{
g_{\tau,\rho} = \left (
\matrix{ {\bf 0} & {\bf -1} \cr
{\bf -1} & {\bf 0} }
\right ),
}
and one can check that there are no projectors which produce invariant
boundary conditions.  Alternatively, one can see that there are no
null $T^2$ subbundles of the doubled geometry.  For such a subbundle,
the $T^2$ fiber must be preserved by the monodromy matrix.  In this
background, the monodromy takes $X^i\to -\tilde X_i$ and $\tilde
X_i\to -X^i$.  So, $X_L$ and $X_R$ are eigenvectors of the monodromy;
for each $i$, the subspaces of definite handedness are monodromy
invariant.  But, the $SO(2,2)$ invariant metric is $ds_L^2 =
2dX^id\tilde X_i = dX^2_L - dX^2_R$, so these subspaces are not null.
Therefore, there is no null $T^2$ subbundle, and no D-brane wrapping
the base once.  On the other hand, since the monodromy squares to the
identity, there do exist D-branes that wrap the base twice.\foot{One
could try to think of these branes as once-wrapped D2-D2 or D1-D3
bound states.  However, since the constituent branes do not exist in
this background, these objects are not strictly speaking bound
states.}

Now consider D0-branes.  It is worthwhile to consider the D-geometry
of this example in two different descriptions, either as a fibration
with monodromy or as an asymmetric orbifold.  As a fibration, we see
that a D0-brane must go around the base twice in order to come back to
itself. Thus, the moduli space of a D0-branes should be a $T^3$ with
one circle of radius $2R$ (the base) and two circles of self-dual
radius $R_{sd}$ (the fiber). From the asymmetric orbifold description,
we would simply start with a base with radius $2R$, and orbifold by
$\tau \to -1/\tau$ and $\rho \to -1/\rho$ as $x
\to x + \pi R$. The moduli space of a D0-brane on this is the same as
the original space, the $T^3$ with radii $2R$, $R_{sd}$, and
$R_{sd}$. The D-geometry is the same in either description.

%%%%%%%%%%%%%%%%%%%%%%%%%%%%%%%%%%
%%%       4. Conclusions       %%%
%%%%%%%%%%%%%%%%%%%%%%%%%%%%%%%%%%

\newsec{Conclusions}
 
We have described simple examples of T-folds with local fiber geometry
$T^2$ and base $S^1$.  These T-folds are interesting in that they
generically have no associated global target space geometry, and are
conveniently described using the doubled torus formalism pursued by
Hull \HullIN.  In this formalism, the background is described by a
$T^{2n}$ fibration over a base manifold $N$, with transition functions
in $O(n,n;\QZ)$.  In each patch, the physical coordinates $X^i$ and
T-dual coordinates $\tX_i$ each span null $T^n$ subspaces determined
by a choice of polarization $GL(n;\QZ)$ in $O(n,n;\QZ)$.  Similarly, a
D-brane wraps a null $T^n$ bundle---the Neumann bundle---over a cycle
of the base. Using this formalism, we have determined the spectrum of
D-branes compatible with each of our T-fold examples.  When any such
\hbox{D-branes} exist, we can associate true global geometries to a
nongeometric background through the classical moduli spaces seen by
these branes.  We have discussed aspects of these \hbox{D-brane}
moduli spaces and described them explicitly to the extent possible.
The sharpest statement can be made when a D-brane is located at point
on the base.  In this case the classical moduli space is the Dirichlet
$T^n$ bundle over $m$-fold cover of the base.

\vskip .3cm

\noindent Three clear avenues for further work were mentioned in the paper:

\vskip .2cm
 
\item{\bf 1.} We would like to better understand quantum dynamics in
the doubled torus formalism.

\vskip .2cm

\item{\bf 2.} Examples on more general base manifolds $N$ should be studied.
This is particularly important for understanding mirror symmetry with
NS-NS flux.

\vskip .2cm

\item{\bf 3.} As Hull points out \HullIN, the doubled torus
description of geometric twisted tori is a real analog of the
description of generalized Calabi-Yau manifolds
\refs{\HitchinUT,\GualtieriThesis}.  The doubling of the torus
corresponds to the use of $TM\oplus T^*M$ for generalized Calabi-Yau
manifolds.  It is only a real analog---in a sense because $M$ is the
fiber rather than the entire manifold.  In Ref.~\DabholkarVE, the
doubling was extended to include the base, at least in the case that
the base is a (doubled) $S^1$.  More generally, in the case that an
even-dimensional underlying structure exists that can play the role of
compactification {\it geometry} \DavidMarks, we would like to see this
connection to generalized complex geometry explored further and made
precise.

\vskip .2cm

\noindent An additional direction to pursue is the following:

\vskip .2cm 

\item{\bf 4.} In this article, 3-form flux corresponded to a $T^2$
fibration over $S^1$ with a particular monodromy.  Even within this
framework, there exist other monodromies corresponding to
``nongeometric" fluxes as studied in \refs{\SheltonCF}.  These are
interesting in their own right, and also give new opportunities for
solving tadpole constraints in flux compactifications.

%%%%%%%%%%%%%%%%%%%%%%%%%%%%%%%%%%%%
%%%       Acknowledgements       %%%
%%%%%%%%%%%%%%%%%%%%%%%%%%%%%%%%%%%%

{\ifnum\lastpenalty>9000\else\bigbreak\fi
\noindent\centerline{\bf Acknowledgements}\nobreak\medskip\nobreak}

We would like to thank Nick Halmagyi, Matthew Headrick, Simeon
Hellerman, Shamit Kachru, John McGreevy, Daniel Ruberman, and Brook
Williams for helpful conversations and comments. B.W. would like to
thank Brook Williams for emotional hospitality and a charming
illustration while this work was being completed. The research of
A.L. is supported in part by NSF grant PHY-0331516, by DOE Grant
No.~DE-FG02-92ER40706, and by a DOE Outstanding Junior Investigator
award.  M.B.S. is supported in part by DOE grant DE-FG02-95ER40893.
The research of B.W. is supported in part by NSF grant PHY-00-96515.

\listrefs
\bye